\documentclass{aa}  

\usepackage{color}
\usepackage{txfonts}
\usepackage{natbib}
\usepackage[breaklinks=true]{hyperref}

\newcommand{\kmps}{km~s\ensuremath{^{-1} }\,}

\newcommand{\Msun}{M\ensuremath{_\odot}\,}

\newcommand{\Oo}{\displaystyle}

\hypersetup{colorlinks,citecolor={blue}}

\begin{document} 

\title{The echo of the bar buckling: \\ Phase-space spirals in Gaia~Data Release 2}
\titlerunning{Echo of the bar buckling}

\author{Sergey Khoperskov\inst{1,2,3}\thanks{sergeykh@mpe.mpg.de}, Paola Di Matteo\inst{2}, Ortwin Gerhard\inst{1}, David Katz\inst{2},\\ Misha Haywood\inst{2}, Françoise Combes\inst{4,5}, Peter Berczik\inst{6,7,8}, Ana Gomez\inst{2}}
\authorrunning{Khoperskov et al. }

\institute{Max Planck Institute for Extraterrestrial Physics, 85741 Garching, Germany \and GEPI, Observatoire de Paris, PSL Universit{\'e}, CNRS,  5 Place Jules Janssen, 92190 Meudon, France \and Institute of Astronomy, Russian Academy of Sciences, 48 Pyatnitskya St., Moscow 119017, Russia \and  Observatoire de Paris, LERMA, CNRS, PSL Univ., UPMC, Sorbonne Univ., F-75014, Paris, France \and College de France, 11 Place Marcelin Berthelot, 75005, Paris, France \and Astronomisches Rechen-Institut, Zentrum fur Astronomie der Universitat Heidelberg, Monchhofstr. 12-14, D-69120 Heidelberg, Germany  \and National Astronomical Observatories of China, Chinese Academy of Sciences, Datun Lu 20A, Chaoyang District, Beijing 100012, China \and Main Astronomical Observatory, National Academy of Sciences of Ukraine, Akademika Zabolotnoho 27, 03680 Kyiv, Ukraine }

 
\abstract{We present a high-resolution numerical study of the phase-space diversity in an isolated Milky Way-type galaxy. Using a single $N$-body simulation~($N\approx 0.14 \times 10^9$) we explore the formation, evolution, and spatial variation of the phase-space spirals similar to those recently discovered by Antoja et al. in the Milky Way disk  with Gaia Data Release 2 (DR2). For the first time in the literature we use a self-consistent $N$-body simulation of an isolated Milky Way-type galaxy to show that the phase-space spirals develop naturally from vertical oscillations driven by the buckling of the stellar bar. Thus, we claim that the physical mechanism standing behind the observed incomplete phase-space mixing process can be internal and not necessarily due to the perturbation induced by a massive satellite. In our model, the bending oscillations propagate outward and  produce axisymmetric variations of the mean vertical coordinate and vertical velocity component of about $100-200$~pc and $1-2$~\kmps, respectively. As a consequence, the phase-space wrapping results in the formation of patterns with various morphologies across the disk, depending on the bar orientation, distance to the galactic center, and time elapsed since the bar buckling.

Once bending waves appear, they are supported for a long time via disk self-gravity. Such vertical oscillations trigger the formation of various time-dependent phase-space spirals in the entire disk. The underlying physical mechanism implies the link between in-plane and vertical motion that leads  directly to phase-space structures whose amplitude and shape are in remarkable agreement with those of the phase-space spirals observed in the Milky Way disk.  In our isolated galaxy simulation, phase-space spirals are still distinguishable  at the solar neighborhood  $3$~Gyr after the buckling phase.  The long-lived character of the phase-space spirals generated by the bar buckling instability cast doubts on the timing argument used so far to get back to the time of the onset of the perturbation: phase-space spirals may have been caused by perturbations originated several gigayearrs ago, and not as recent as suggested so far. }


\maketitle

\section{Introduction}\label{sec::intro}

After two successful data releases, Gaia already provided to the community an unprecedented amount of high-quality knowledge about kinematics and spatial structures of the Milky Way disk and its surroundings~\citep{2016A&A...595A...2G,2018A&A...616A...1G}. In particular, the second data release (DR2) contains the full six-dimensional phase-space information for $7.2$~million stars, which made possible several discoveries about kinematics and the spatial structure of our Galaxy~\citep{2018A&A...616A..11G}. It has become abundantly clear that nonequilibrium processes in the Milky Way such as satellite interactions, bar dynamics, and spiral structure effects have shaped the present-day structure and kinematics of the Milky Way significantly.  One of the most striking new features discovered is the spiral-like distributions in the $z-v_z$ plane, which suggest that the Milky Way disk is locally out of dynamical equilibrium. In particular, \cite{2018Natur.561..360A} found spirals in the distribution of radial and azimuthal velocities in $z-v_z$ plane for stars that lie in a small volume around the Sun~($<0.1$~kpc). By combining Gaia~DR2 with spectroscopic surveys such as GALAH~\citep{2018arXiv180902658B}  and LAMOST~\citep{2018ApJ...865L..19T}, this result has been quickly confirmed to hold also in a more extended volume of the Milky Way disk. 

\begin{figure*}
\centering
\includegraphics[clip=true,width=1\hsize]{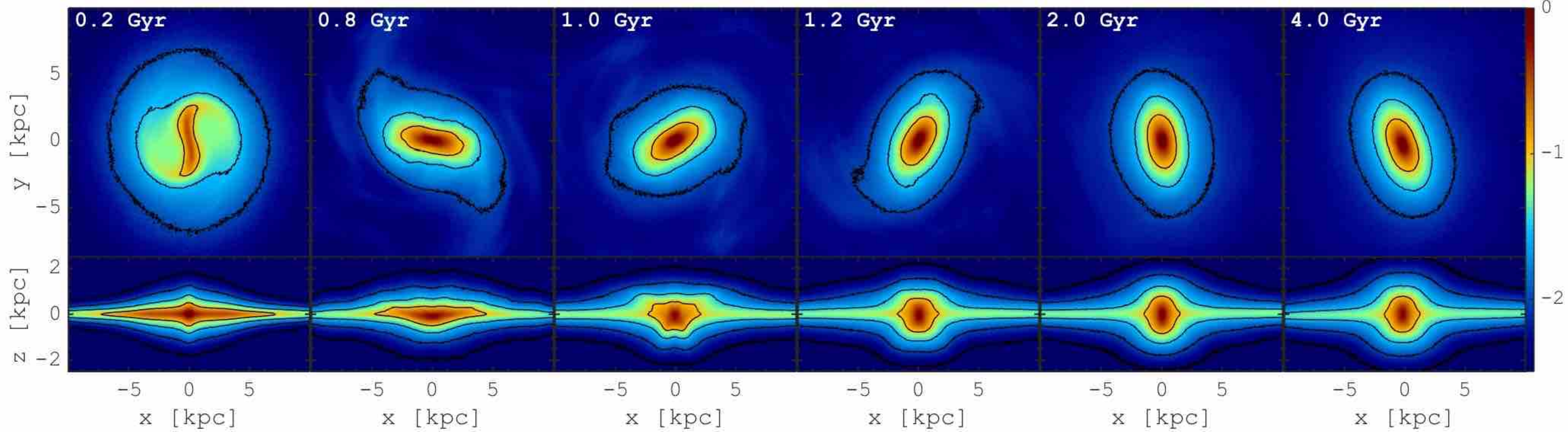}
\caption{Face-on and edge-on projected density evolution (in log-scale) of the galactic disk. Buckling phase of the bar is depicted in second and third frames. We note that beyond the bar radius the disk density distribution stays axisymmetric during the entire simulation.}\label{fig::fig1}
\end{figure*}

Although we already know some signatures of the nonequilibrium state of the Milky Way disk in the solar suburb~\citep[see, e.g., ][]{2011MNRAS.412.2026S, 2012ApJ...750L..41W, 2013MNRAS.436..101W,2015ApJ...800...83B,2013ApJ...777L...5C,2017ApJ...847..123P,2018MNRAS.475.2679C}, the exact mechanism driving the phase-space $z-v_z$ spirals is still poorly constrained. For instance, \cite{2012ApJ...750L..41W}  proposed that the corrugated velocity patterns in Milky Way-type galaxies can be generated by a passing satellite or  dark matter subhalos~\citep[see also][]{2015MNRAS.446.1000F,2016ApJ...823....4D}. In particular, a slow moving satellite induces a bending-mode perturbation. However, the oscillations decay rapidly in time because of phase mixing and Landau damping, which lead to disk heating.  In such a scenario,  the velocity patterns seen in the observations should indicate a very recent tidal interaction~\citep[see also][]{2014MNRAS.440.1971W}. In such a framework \cite{2018Natur.561..360A} interpreted the phase-space spirals as a signature of continuous ongoing mixing in the vertical direction, possibly excited by a recent encounter with the Sagittarius dwarf galaxy.  By making use of test particle simulations, these authors estimated that the vertical phase mixing event started about $300-900$~Myr ago. This suggestion has been confirmed qualitatively in a low-resolution $N$-body simulation of the interaction of Sagittarius with the Galactic disk~\citep{2018arXiv180800451L}. More detailed analysis has been carried out by ~\cite{2018arXiv180902658B} who show that to produce a realistic phase-space spiral,  the mass of perturber should be higher than $10^{10}$~\Msun. Such value is in agreement with another recent dynamical study of the spirals in the $(z,v_z)$-plane by~\cite{2018MNRAS.481.1501B}, who noted, however, that the main weakness of such an external perturbation scenario is in the high values of mass and passage duration, which are needed to obtain a realistic model of the phase-space spirals.

\begin{figure}
\centering
\includegraphics[clip=true,width=0.9\hsize]{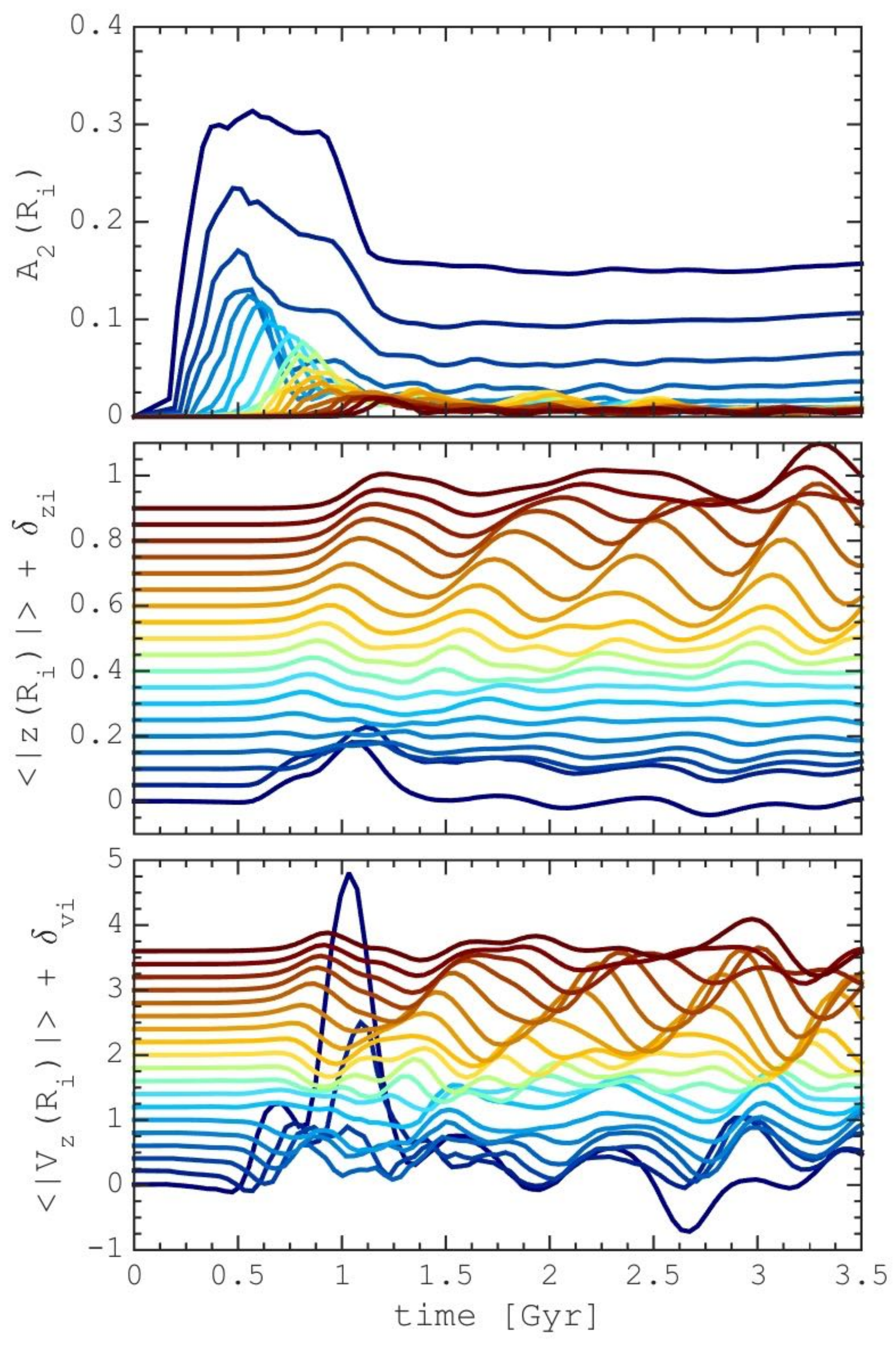}
\caption{Evolution of the bar strength $A_2$~(top), median vertical coordinate of particles~(middle), and mean vertical velocity~(bottom). All frames show the evolution of parameters at different radii in a circular annulus of $0.5$~kpc width: from blue in the center to brown lines at the outskirts. To improve the visibility of the $\langle z \rangle$,$\langle v_z \rangle $ time-dependent variations we shift each curve relative to the previous by a constant value of $\delta_{\rm vi} = 0.2$ and $\delta_{\rm ri} = 0.05$ for vertical velocity and vertical coordinate respectively.  The end of the bar buckling phase corresponds to $\approx 1$~Gyr. }\label{fig::fig2}
\end{figure}

Another mechanism proposed to explain the continuous phase-space mixing is the spontaneous generation of bending waves in isolated disk galaxies~\citep{2018arXiv180711516D}. By making use of  test particles simulations and $N$-body models, these authors claimed that self-gravity is important to coupling of the in-plane and vertical motions that lead to the formation of the spirals detected by~\cite{2018Natur.561..360A}. We note however that a limitation of  their model is the use of an ad hoc imposed bending perturbation of velocity of about $30$~\kmps. In such a picture, the properties of phase-space spirals and their duration directly depend on the strength of the imposed perturbation and hence there is no difference between the isolated model with an ad hoc vertical perturbation and the excitation of phase-space spirals due to a passing Sagittarius dwarf galaxy. 

Internal processes, external tides, and local galaxy disk perturbations are the main physical factors that could be responsible for the observed phase-space mixing process in the solar neighborhood, but we still do not know the dominant cause of the observed deviations from equilibrium state. In this Letter, we want to study and discuss an alternative scenario for the formation of the  phase-space spirals found in Gaia DR2, related to internal mechanisms, and not to external perturbations, by showing in particular that these spirals may be the last vestiges of the bar buckling instability. This study is performed using an extreme high-resolution three-dimensional $N$-body simulation. The outline of the Letter is as follows: in Sect.~\ref{sec::sec2}, we describe our model; in Sect.~\ref{sec::sec3}, we examine the phase-space structure of the simulated Milky Way-type galaxy, which develops a bar, that then buckles and forms a boxy/peanut bulge; and finally, in Sect.~\ref{sec::sec4}, we summarize the conclusions that can be drawn from our study.

\section{Model}\label{sec::sec2}
In order to investigate the phase-space mixing in the Milky Way type galaxy, we performed a single, extremely high-resolution, $N$-body simulation of disk galaxy evolution embedded into a live dark matter halo. We adopted the initial conditions from a recent study of the bar/bulge regions of the Milky Way by \cite{2017A&A...607L...4F} and details are presented in Appendix~\ref{app1}. 

For a deeper understanding of the evolution in the phase-space distribution of disk stars, we require simulations that are capable of resolving the dynamics down to few tens of parsecs scale; this resolution is much higher than most of the current Milky Way-like galaxies simulations. In order to get inside the rich substructures in the phase space, we used $10^8$ particles to model the stellar disk component and $41 451 200$~particles to model the live dark matter halo. To our knowledge, this is one of the highest resolution, $N$-body simulations of a disk galaxy.

\begin{figure*}
\centering
\includegraphics[clip=true,width=0.9\hsize]{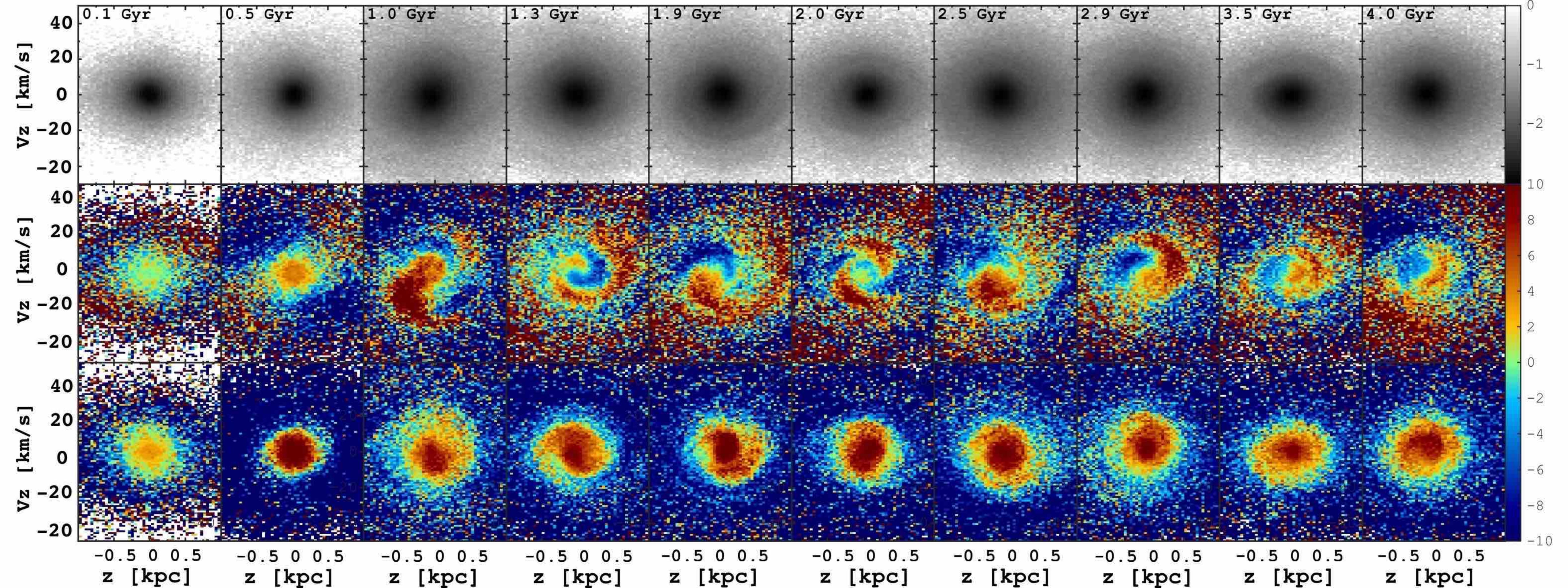}
\caption{Evolution of phase-space spiral in $z-v_z$ plane for a solar neighborhood-like region in the disk, at $R=8$~kpc. At all times~(except the first frame) the bar is rotated by $30$ degrees relative to the direction on the galactic center. {\it Top row:} stellar density distribution $\Oo \log( \rho(z,v_z)/\max(\rho(z,v_z)) )$, {\it middle row:} the mean radial velocity, and {\it bottom row:} the mean residual azimuthal velocity $v_\varphi(z,v_z)- \langle v_\varphi(z,v_z) \rangle$. }\label{fig::fig3}
\end{figure*}

\section{Results}\label{sec::sec3}
The global evolution of our simulated galaxy is in good agreement with many previous $N$-body models in which bars are usually weakened by the buckling instability~\citep[see, e.g.,][]{1991Natur.352..411R,1990A&A...233...82C,2004ApJ...604L..93D,2005MNRAS.358.1477A}. In our simulation, the initially axisymmetric disk develops a prominent bar in about two-three rotation periods. Between about $0.5$ and $1$ Gyr, the bar becomes vertically unstable and buckles, breaking the symmetry with respect to the equatorial disk midplane~(see Fig.~\ref{fig::fig1}). After $\approx 1.2$~Gyr, the bar profile across the disk plane becomes again symmetric, but the inner disk region now has a larger vertical thickness, and its isodensity-contours show a clear boxy morphology.  

We may state that, so far, in relation to the study of the bar buckling process,attention has mostly been drawn to collective phenomena associated with the formation of bulges, while its possible impact  on the outer disk regions -- that is essentially outside the bar corotation -- has not received particular attention. The bending instability has been associated with nonlinear processes during the bar evolution. The growth of the bar involves  stars with low-velocity dispersion from the outer to inner regions, where the disk is already significantly hotter. Such a kinematical pressure anisotropy appears to drive the fire-horse instability across the bar~\citep{1991Natur.352..411R}, which in turn leads to the thickening of the disk, as a way  to remove the source of instability in the disk. In addition to this, we find that the process of the vertical bar instability~(see Fig.~\ref{fig::fig1}) is also reflected in the kinematical properties of stars in the outer~(relative to the bar) disk. 

As follows from our simulation, indeed the development of the boxy/peanut structure also affects   the disk areas outside the bar region. To demonstrate this in Fig.~\ref{fig::fig2} we present the evolution of the bar strength~(Fourier harmonics $A_2$), median vertical offset $\langle z \rangle$ and median vertical velocity $\langle v_z \rangle$ measured for stars at different galactocentric radii from $1$ to $10$~kpc. During the early linear phase of the bar growth~($<0.5$~Gyr) our model is stable against the bending instability at all radii. However, once the disk starts to buckle and the bar strength decreases, bending waves appear initially in the central part of the disk, the bar region, forming bell-like structures, which later propagate farther out toward the galactic outskirts. After a relatively short vertical impulse in which the edge of the buckling bar acts as a source of bending waves, quasi-stationary ring-like structures are established over the entire outer galaxy. In particular, a symmetric $m=0$ perturbation moves outward, and its amplitude weakly increases in the outer parts of the disk. It takes roughly $0.5$~Gyr for perturbations to reach the outer disk part~($>10$~kpc). Such a bending wave propagation process is very much similar to those found in a number of various theoretical and numerical works about the excitation of internally driven bending waves in the galactic disks~\citep[see, e.g.,][]{1969ApJ...155..747H,1971Ap&SS..14...52K,1994ApJ...425..551M,2010AN....331..731K,2013MNRAS.434.2373R, 2017A&A...597A.103K, 2017MNRAS.472.2751C}.

The consequences of the bar buckling can persist over a significant time -- more than 10 rotational periods at the solar radius -- and it can have observational counterparts in the $(z,v_z)$ phase-space distribution. Indeed, independent of the mechanism, the bending waves propagating across the disk suppress the mixing process in the phase space. It is thus natural to investigate the kind of  features left by this mechanism in the  $z-v_z$ plane. In order to compare our simulation with the results by ~\cite{2018Natur.561..360A}, in Fig.~\ref{fig::fig3} we present the evolution of the phase-space mixing patterns in the disk at $R=8$~kpc from the center in a small arc of $15^\circ$ of $1$ kpc radial width oriented with $30^\circ$ relative to the bar. This region contains of about $547856$ to $624231$ particles at different times. At early times, before the perturbation from the buckling bar reaches the solar radius, $(z,v_z)$ phase-space structure is highly symmetric in the innermost part~(low $|\langle z \rangle|$ and $|\langle v_z \rangle|$), and only a weak quadrupole pattern dominates in the outer region of the distribution. Such a picture is known as a tilt of the velocity ellipsoid~\citep[see, e.g., Section 5.1 in][and references therein]{2018arXiv180902658B}, which is the manifestation of a fully mixed galactic disk. Once the vertical perturbations arrive at the outer disk regions~($T\approx0.5$~Gyr), in the $v_r(z,v_z)$ distribution, we start to observe an asymmetric two-arm spiral structure in $(z,v_z)$-plane, which corresponds to a transition phase between a tilted ellipse and a single arm spiral. Figure~\ref{fig::fig3} demonstrates a clear similarity to the Gaia-like phase-space spiral at $T=1.3$~Gyr. The spiral structure is more clearly seen in the distribution of radial velocity  $(z,v_z)$ space; this structure is less evident in azimuthal velocity, but still distinguishable in density distribution. Thus we find that internally driven bending waves trigger a clear correlation between in-plane and vertical motions that lead to spiral structures that are in good agreement to the phase-space spirals observed at the solar neighborhood.

Surprisingly, the phase-space spiral pattern is still visible in a solar neighborhood-like region at later times~($T>1.5$~Gyr) after the end of the boxy/peanut bulge formation. The reason for this is that in our model the bar-driven bending waves do not dissipate even after the initial source~(buckling of the bar) disappears and the weave-like perturbations persist in the disk for tens of rotational periods~(up to $4$~Gyr). The bending waves~(see Fig.~\ref{fig::fig2}) are not kinematical because they do not dissipate on the dynamical timescale. This has been  predicted  by~\citep[][]{2018arXiv180711516D}, who showed that self-gravity is crucial for the persistence of long-lived spontaneous bending waves in disks. Hence disk self-gravity is essential for the support of bending waves triggered by the bar buckling. That is why, even a very long time~($\approx3$~Gyr) after the buckling of the bar -- $z-v_z$ spiral patterns are clearly seen. However, the morphology of spirals is very much time-dependent, and the outer part of the distribution always shows the existence of a quadrupole structure, which is visible as a top right to bottom left reddish asymmetry in the radial velocity distribution $v_r(z,v_z)$~(see Fig.~\ref{fig::fig3}).

\begin{figure*}
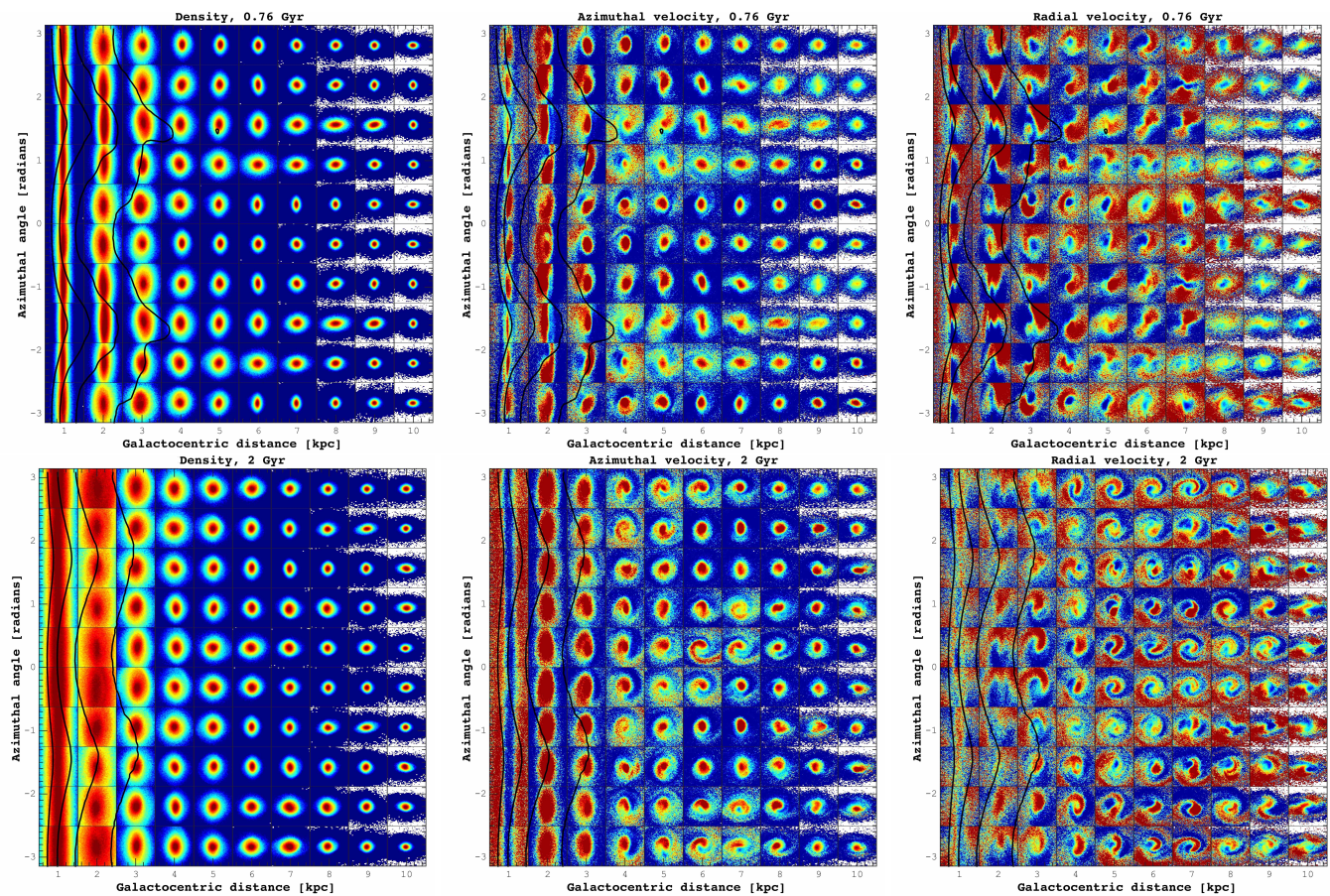

\begin{center}
\includegraphics[clip=true,width=0.32\hsize]{figs/spiral_spiral_run6M_0019.pdf}\includegraphics[clip=true,width=0.32\hsize]{figs/spiral_up_run6M_0019.pdf}\includegraphics[clip=true,width=0.32\hsize]{figs/spiral_ur_run6M_0019.pdf}\\
\includegraphics[clip=true,width=0.32\hsize]{figs/spiral_spiral_run6M2_0050.pdf}
\includegraphics[clip=true,width=0.32\hsize]{figs/spiral_up_run6M2_0050.pdf}
\includegraphics[clip=true,width=0.32\hsize]{figs/spiral_ur_run6M2_0050.pdf}
\end{center}
\caption{Spatial variations of the phase-space patterns in simulated galaxy at $0.76$~Gyr~(top) and at $2$~Gyr~(bottom). Each frame corresponds to number density~(left), mean azimuthal~(middle), and radial velocities~(right) in $z-v_z$ plane for stars selected in small arcs $|R-R_i|<0.5$~kpc and $|\phi - \phi_i|<7.5^\circ$, where cylindrical coordinates $(R_i,\phi_i)$ are indicated on global axes. Each frame axes size is $\rm [-1;1]\, kpc \times [-50;50]\,\kmps $. Total galaxy stellar surface density contours are over-plotted by black lines, which clearly show the position of the bar in galactic cylindrical coordinates. Number of particles varies from $\approx 4\times 10^6$ in the innermost frames to $\approx 0.3\times 10^6$ in the outermost frames.}\label{fig::fig4}
\end{figure*}

The large number of particles in our simulation enable us to derive a global view of the phase-space disk structure. This allows us to understand the uniqueness of the phase-space structure recently discovered in the extended solar vicinity. To do this, in Fig.~\ref{fig::fig4},~\ref{fig::fig5}, we plot the distributions of stellar density, radial, and azimuthal velocities in the $z-v_z$ plane for different parts of the disk at different times: $0.76$~Gyr~(during the bar buckling) and at $2$~Gyr~(when the boxy-bulge is formed). Figure \ref{fig::fig5} shows the same distributions but at very late times $4$~Gyr. These plots show a clear evidence of ongoing phase-space mixing occurring across the entire galaxy. In particular, at early times, when the bending waves are just triggered by the bar, the inner disk~($R\leq5$) shows the presence of one-arm spiral structures. While the outer disk~($R>5$) is not vertically perturbed yet and various symmetric quadrupole-like structures for radial velocity distribution in $z-v_z$ plane persist there.  Thanks to the homogeneous outer disk~(beyond the bar size, see Fig.\ref{fig::fig1}), we obtain clear periodic variations of the phase-space patterns depending on the bar position angle~(see black contours in Figure~\ref{fig::fig4},~\ref{fig::fig5}). Thus, we find that the phase-space structure of disk stars in an isolated, barred Milky Way-type galaxy varies across the disk and the spiral pattern in $z-v_z$ plane is nonsteady, but long-lived. This is because the disk self-gravity is important to support and enhance the bending waves beyond the bar radius, even few gigayear after the end of the bar buckling phase.  

\section{Conclusions}\label{sec::sec4}

In this paper, which is part of a series exploring various chemo-kinematical features found by Gaia DR2 and large-scale spectroscopic surveys~(GALAH, LAMOST), we examine the rich phase-space structure of stellar populations in the Milky Way-type disk. To this aim, we present a model of the natural formation of the spiral patterns in $z-v_z$-plane in an isolated galaxy with a boxy/peanut bulge, such as that of the Milky Way, where the mechanism driving the vertical oscillations is the buckling instability of the bar. The existence of long-lived vertical oscillations in stellar disk has been explored by numerous $N$-body simulations, but in this work we demonstrate a clear impact of the bar buckling phase on the generation of bending waves propagating across the disk. As a consequence, coherent vertical and in-plane motions are both responsible for generating various phase-space patterns in the galactic disk.

Our model is the first self-consistent, isolated galaxy model that is successful in reproducing the spiral structures in the phase-space discovered by \cite{2018Natur.561..360A} and it reveals a rich phase-space structure over the entire Milky Way-type galaxy disk. We have been able to show that the phase-space spirals are related to a nonequilibrium state of the Milky Way disk, and by means of comprehensive analysis of a high-resolution $N$-body simulation, we demonstrate clearly that it is hard to complete the phase-space relaxation even in an isolated galaxy. We confirm the conclusions of the studies by~\cite{2014MNRAS.440.1971W} and \cite{2018arXiv180711516D} that self-gravitating disks are able to support long-lived vertical oscillations even when  the source of bending instability  is gone. That is why in our pure $N$-body simulation the bending waves do not dissipate quickly after the end of the bar buckling.

In the understanding of the origin of the phase-space spirals, we are thus left with two possible mechanisms: an external, recent perturbation by a massive satellite, as suggested by \cite{2018Natur.561..360A}, \cite{2018MNRAS.481.1501B}, and \cite{2018arXiv180902658B}, and an internal mechanism, related to the secular evolution of the disk, as shown in this paper. The main limitations of the accretion scenario are currently related to the high mass of the perturber and the recent onset of the perturbation. As discussed by \cite{2018MNRAS.481.1501B}, indeed, the perturbing satellite would need to be massive and the duration of the effect brief in time. A relatively high mass for the perturber is also found by \cite{2018arXiv180902658B}, at the level of $3x10^{10}M_\odot$. Perhaps the mass argument is less important though, since there have been suggestions that the progenitor of Sagittarius dwarf galaxy could have been more massive than initially envisaged~\cite[see][$10^{11}M_\odot$]{2017MNRAS.464..794G}, but  still this is the original mass of the satellite and not the mass after several Gyrs of stripping. The internal scenario does not suffer of any of these two limitations. In our model, we have not fine-tuned either the strengths of the bar and of the boxy/peanut bulge or the timescale of the buckling instability. Interestingly, we find that the timing argument can be relaxed, that is we do not need the perturbation to be as recent as predicted in the case of the satellite accretion. However, it will be necessary to address in future works the upper limit of this timescale in the case in which the phase-space spirals originate from the buckling instability. Could the buckling have occurred about $8$~Gyr ago at the time when bars are expected to appear in galaxies of the size of the Milky Way~\citep[see][]{2008ApJ...675.1141S,  2012ApJ...757...60K}, or do we need a more recent or maybe recurrent~\citep[see][]{2006ApJ...637..214M} buckling to explain the observed phase-space patterns? Also, the role of a dissipative component in counteracting the secular heating of the disk, the role of spiral
asymmetries,  and the Gaia selection function will need to be taken into account in future models.  Despite these limitations, our work shows once more that the role of the bar, in our Galaxy, may have been fundamental also in generating of long-lasting phase-space patterns and incomplete mixing.

\begin{acknowledgements}
We thank the anonymous referee for her/his valuable comments that helped to improve this paper. This work was granted access to the HPC resources of CINES under the allocation 2017-040507 (PI : P. Di Matteo) made by GENCI. This work has been supported by the ANR (Agence Nationale de la Recherche) through the MOD4Gaia project (ANR-15-CE31-0007, P.I.: P. Di Matteo). Numerical simulations are carried out partially with support by the Russian Foundation for Basic Research (16-32-60043, 16-02-00649) and using the equipment of the shared research facilities of HPC computing resources at Lomonosov Moscow State University supported by the project RFMEFI62117X0011. PB acknowledge the support by Chinese Academy of Sciences through the Silk Road Project at NAOC, through the ``Qianren'' special foreign experts program and also under the President's International Fellowship for Visiting Scientists program of CAS. PB acknowledges the partial financial support of this work by the Deutsche Forschungsgemeinschaft through Collaborative Research Center (SFB 881) "The Milky Way System" (subproject Z2) at the Ruprecht-Karls-Universitat Heidelberg. PB acknowledge the support of the Volkswagen Foundation under the Trilateral Partnerships grant No. 90411 and the special support by the NASU under the Main Astronomical Observatory GRID/GPU computing cluster project.

\end{acknowledgements}

\begin{appendix}
\section{Model description}\label{app1}
\begin{figure}
\begin{center}
\includegraphics[clip=true,width=1\hsize]{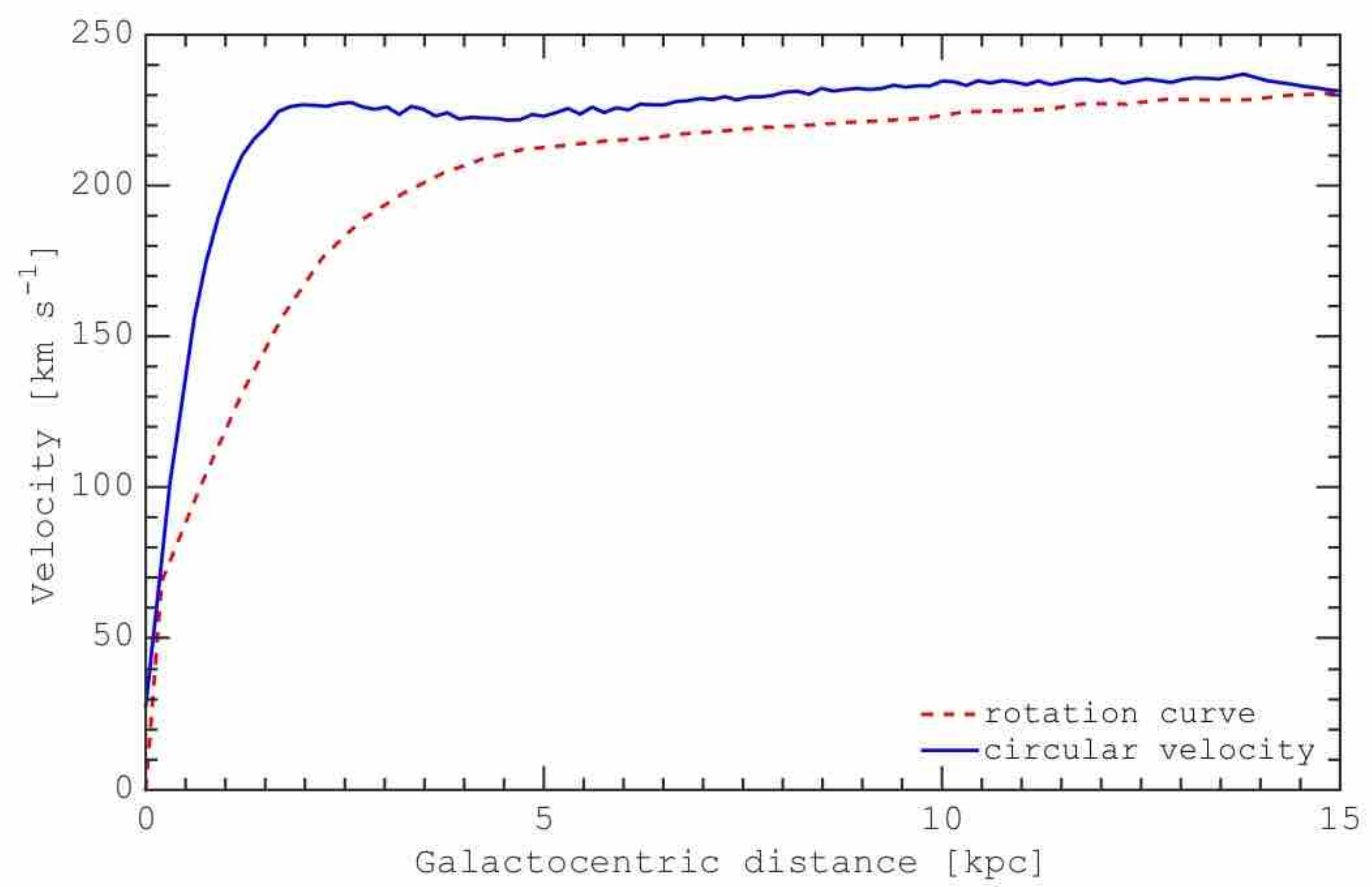}
\end{center}
\caption{Circular velocity (blue curve) of the model, together with the rotation curve (red).}\label{fig::fig0}
\end{figure}
In this work we employ a multicomponent model for a galaxy consisting of three cospatial disk populations with different parameters. Each disk component was represented by a Miyamoto-Nagai density profile~\citep{1975PASJ...27..533M} that has a characteristic scale length of $2$, $2,$ and $4.8$ kpc, vertical thicknesses of $0.15$, $0.3,$ and $0.6$~kpc and masses of $1.86$, $2.57,$ and $4.21 \times 10^{10}$~\Msun , respectively. Our simulation also includes a live dark matter halo whose density distribution follows a Plummer sphere~\citep{1911MNRAS..71..460P}, which has a total mass of $3.81\times 10^{11}$~\Msun and a radius of $21$~kpc. The choice of parameters leads to a galaxy mass model with a circular velocity of $\approx220$~\kmps and Toomre stability parameter $Q_T\approx1.5$ at $8$~kpc. The initial setup has been generated using the iterative method by~\citet{2009MNRAS.392..904R}.

To avoid small-scale bends due to initial artificial  nonaxisymmetries, we forced a proper symmetry for the initial positions and velocities of all particles in the simulation~\citep[e.g., similar to][]{2014MNRAS.443L...1D}. As a consequence, the stellar disk is highly symmetric about the midplane. This model,  where the thick disk is massive and centrally concentrated, can reproduce the morphology of the metal-rich and metal-poor stellar populations in the Milky Way bulge, as well as the mean metallicity and $\rm [\alpha/Fe]$ maps as obtained from the APOGEE data~\citep{2018A&A...616A.180F}. Hence we emphasize that our model is meant to be representative of the Milky Way disk, its stellar populations, and global kinematics.

For the $N$-body system integration, we used our parallel version of the TREE-GRAPE code~\citep[][Section 4.1]{2005PASJ...57.1009F} with multithread usage under the SSE and AVX instructions. We also ported the original GRAPE-based~(GRAvity PipEline) tree code to few different hardware platforms, including the CPU and the recent 
IA Graphics Processing Unit platform using the Compute Unified Device Architecture. In recent years we already used and extensively tested our hardware-accelerator-based gravity calculation routine in several galaxy dynamics studies where we obtained accurate results with a good performance~\citep{2005KFNT...21..288S, 2008AN....329.1029B,2010A&A...514A..47P, 2011A&A...525A..99P}. In the simulation we adopted the standard opening angle $\theta=0.5$ and a gravitational softening parameter equal to $10$~pc. For the time integration, we used a leapfrog integrator with a fixed step size of $0.1$~Myr~\citep{2014JPhCS.510a2011K}.

\section{Additional plot}\label{app2}

\begin{figure*}
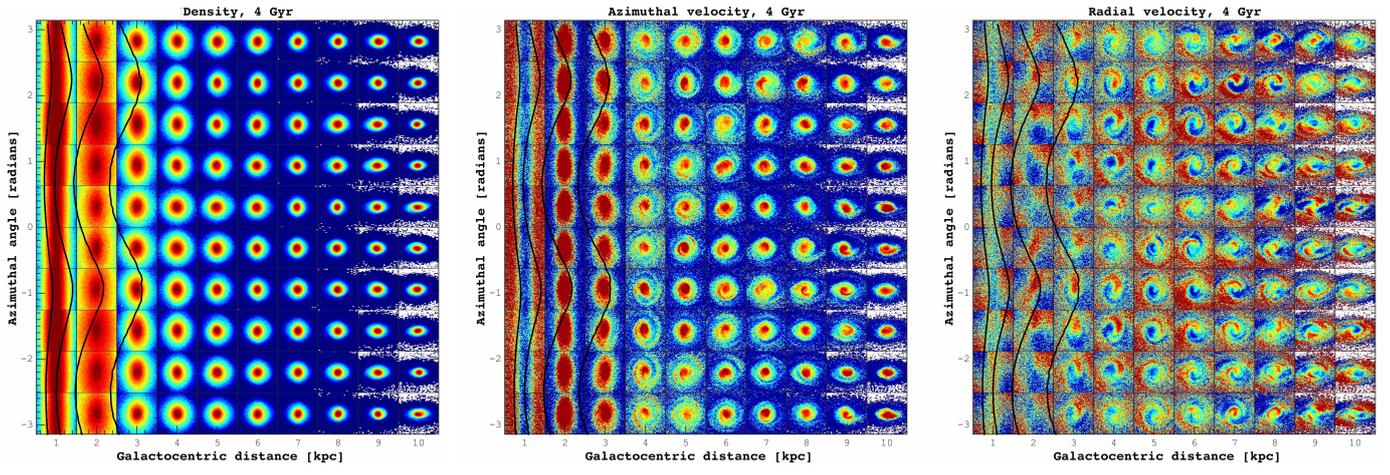

\begin{center}
\includegraphics[clip=true,width=0.33\hsize]{figs/spiral_spiral_run6M2_0100.pdf}
\includegraphics[clip=true,width=0.33\hsize]{figs/spiral_up_run6M2_0100.pdf}
\includegraphics[clip=true,width=0.33\hsize]{figs/spiral_ur_run6M2_0100.pdf}
\end{center}
\caption{Same as in Fig.~\ref{fig::fig4}, but for latter time of $4$~Gyr.}\label{fig::fig5}
\end{figure*}

\end{appendix}

\bibliographystyle{aa}
\bibliography{references}

\end{document}